\documentclass[a4j,10pt,onecolumn,oneside,notitlepage,final]{article}
\usepackage{alltt}
\usepackage{graphicx} 
\usepackage{amsmath}
\usepackage{amssymb}
\usepackage{amscd}
\usepackage{graphicx}

\title{Observables Generalizing Positive Operator Valued Measures}

\author{Irina Basieva and Andrei Khrennikov\\
International Center for Mathematical Modelling\\
in Physics and Cognitive Sciences\\ Linnaeus University, V\"axj\"o,
S-35195, Sweden}

\begin{document}
\maketitle

\begin{abstract}
We discuss a generalization of POVM which is used in quantum-like modeling of mental processing. 
\end{abstract}

\section{Introduction}

Recently quantum information was actively used in applications to cognitive science, psychology, 
genetics, see, e.g., \cite{AAA}-\cite{AC} and references herein.  As was pointed out in \cite{AC} , although many mental models cannot be described by the 
classical probabilistic model (Kolmogorov, 1933; at least multi-Kolmogorovian framework has to be explored), the quantum 
probabilistic formalism also may be too special to cover all sorts of mental probabilistic nonclassicality. In \cite{AAA} it was shown that
there exist statistical data from cognitive psychology which cannot be represented with the aid of standard quantum 
observables given by Hermitian operators (the Dirac-von Neumann quantum formalism). 
It seems that non-Hermitian operators has to be involved in operational representation of mental observables. 
In \cite{AAA} a {\it quantum-like representation algorithm} (QLRA) was  elaborated.  
It can be applied to produce operator representation of 
observables (of any origin: physical, mental, biological), but operators are not always Hermitian. This is a consequence
of the experimental fact that, for statistical data collected in cognitive psychology, the matrices of transition 
probabilities  are not always doubly stochastic, non-doubly stochastic matrices can arise as well. And we know that
the matrix of transition probabilities for two quantum observables (with nondegenerate spectra) given by Hermitian operators   
is always doubly stochastic.
Moreover, the author  of \cite{AAA} was not able
to describe aforementioned statistical data even by using positive operator valued measures (POVMs) which represent generalized  
quantum observables in quantum information theory. In this note we describe the class of observables which are produced by  
QLRA and study its connection with the class of ``conventional generalized observables'', POVMs. We also study the question whether the statistical data described by ``new generalized observables'' can also  be described by POVMs. We show that the class of new generalized observables is larger than the class of POVMs. 

\section{New generalization of quantum mechanical  formalism}

Let us consider a finite dimensional Hilbert space $H.$ Let ${\cal
E}=\{e_j\}_{j=1}^n$ be an orthonormal basis:
\begin{equation} \label{BBBB}
\psi=\sum_j c_j e_j, c_j=c_j(\psi) \in {\bf C}.
\end{equation}
 Each ${\cal E}$
generates a class of (conventional) quantum observables,
self-adjoint operators:
\begin{equation} \label{SA} \hat{a} \psi=\sum_j y_j c_j(\psi) e_j,
\end{equation}
where $X_a=\{y_1, ..., y_n\}, y_j \in {\bf R},y_j\not= y_i $ is
the range of values of $a$ (so we start with consideration of
observables with nondegenerate spectra).

Let now ${\cal E}=\{e_j\}_{j=1}^n$ be an arbitrary basis (thus in
general $\langle e_j, e_i \rangle \not=0, i \not= j)$ consisting
of normalized vectors, i.e., $\langle e_j, e_j
\rangle=1.$

 We generalize the Dirac-von Neumann formalism by considering
observables (\ref{SA}) for an arbitrary ${\cal E}.$ We also
consider an arbitrary nonzero vector of $H$ as a pure quantum
state. We postulate (by generalizing Born's postulate):
\begin{equation} \label{SA1} P_\psi(a=y_j)= \frac{\vert
c_j(\psi)\vert^2}{\sum_j\vert c_j(\psi)\vert^2},
\end{equation}
where the coefficients $c_j(\psi)$ are given by the expansion
(\ref{BBBB}).

 If ${\cal E}$ is an orthonormal basis, then
$c_j(\psi)=\langle \psi, e_j \rangle, \sum_j\vert
c_j(\psi)\vert^2= \Vert \psi \Vert^2$ and for a normalized vector
$\psi,$ we obtain the ordinary Born's rule.

Our generalization of the Dirac-von Neumann formalism is also very
close to another well known (and very popular in quantum information theory)
generalization of the class of quantum observables, namely, to the
formalism of POVMs. To proceed in this
way, we introduce projectors on the basis vectors: $\pi_j
\psi=c_j(\psi) e_j.$ We remark that $\pi_j^2= \pi_j,$ but in
general $\pi_j^*\not= \pi_j.$ We have: $\vert c_j(\psi)\vert^2=
\langle \pi_j \psi, \pi_j \psi\rangle= \langle M_j \psi,
\psi\rangle,$ where $M_j= \pi_j^* \pi_j.$ We remark that each
$M_j$ is self-adjoint and, moreover, positively defined. We also
set $M=\sum_j M_j.$ Then our generalization of Born's rule can be
written as: \begin{equation} \label{SA2} P_\psi(a=y_j)=
\frac{\langle M_j \psi, \psi\rangle }{\langle M \psi,
\psi\rangle}= \frac{\rm{Tr}\; \rho_\psi M_j}{\rm{Tr}\; \rho_\psi
M},
\end{equation}
where $\rho_\psi= \vert \psi\rangle  \langle\psi\vert.$ We remark that, for an
arbitrary nonzero $\psi,$ the operator $\rho_\psi\geq 0.$

Now we generalize the conventional notion of the density operator,
by considering any nonzero $\rho \geq 0$ as a generalized density
operator. The corresponding generalization of
Born's postulate has the following form:
\begin{equation} \label{SA3} P_\psi(a=y_j)= \frac{\rm{Tr}\; \rho\;
M_j}{\rm{Tr}\; \rho \;M}.
\end{equation}
The only difference from the POVM formalism is that the operator
$M\not=I$ (the unit operator).

We remark that $\langle M\psi, \psi \rangle= \sum_j \vert
c_j(\psi) \vert^2 \not= 0, \psi \not=0.$ Thus (we are in the
finite dimensional case) the inverse operator $M^{-1}$ is well
defined.

We now proceed with our formalization and consider an arbitrary
(separable) Hilbert space $H.$

{\bf Definition 1.} {\it A generalized quantum state is
represented by an arbitrary trace class nonnegative (nonzero)
operator $\rho: \rho \geq 0, 0 < \rm{Tr} \rho < \infty.$}

{\bf Definition 2.} {\it A generalized quantum observable is
represented by an arbitrary (so in general non normalized)
positive operator valued measure $E$ on a measurable space
$(X,{\cal F})$ such that $E(X)>0.$}

Thus, for a generalized quantum observable $E,$   we have:

1). $E(B)\geq 0,$ for any set $B\in {\cal F},$ and $E(X)>0;$

2). $E(B)^*=E(B),$ for any set $B\in {\cal F};$

3). $E(\cup_{j=1}^n B_j)=\sum_{j=1}^n E(B_j)$ for all disjoint
sequences $\{B_j\}$ in ${\cal F}.$

\medskip

{\bf Generalized Born's rule:} Let $\rho$ and $E$ be generalized
quantum state and observable, respectively. Then the probability
to find the result $x$ of the $E$-measurement  in a measurable set
$B$ (for an ensemble represented by $\rho$) is given by
\begin{equation}
\label{QST} P_\rho(x \in B)= \frac{\rm{Tr} \rho \;E(B)}{\rm{Tr}
\rho \;E(X)}.
\end{equation}

We remark that $\rm{Tr} \rho \;E(X)>0.$ To prove this, we consider
the spectral expansion of  the trace class operator $\rho= \sum_j
q_j \psi_j\otimes\psi_j.$ Here at least one $q_j>0.$ Then $\rm{Tr}
\rho \;E(X)=\sum_j q_j \langle E(X)\psi_j, \psi_j\rangle>0.$

We remark that probabilities with respect to non normalized (generalized) quantum state
$\rho,$ i.e., $\rm{Tr} \rho \not=1,$ always can be rewritten as probabilities with respect 
to the corresponding normalized (i.e., usual) quantum state. Set 
\begin{equation}
\label{NR}
\rho^\prime= \rho/  \rm{Tr} \rho.
\end{equation}
Then we can scale the nominator and denominator in (\ref{QST}) by diving  by
$ \rm{Tr} \rho$ and obtain:
\begin{equation}
\label{QST1} P_\rho(x \in B)=  P_{\rho^\prime}(x \in B)= \frac{\rm{Tr} \rho^\prime \;E(B)}{\rm{Tr}
\rho^\prime \;E(X)}.
\end{equation}
Hence, on the level of probabilities the magnitude of the trace of a generalized quantum state does not play 
any role. Therefore we can restrict our generalization of quantum formalism just to observables and operate with 
standard quantum states.\footnote{However, non normalized generalizations of quantum states can appear quite naturally
in some models, e.g., in prequantum classical statistical field theory, see, e.g., \cite{AC1}. Here, non normalized generalizations of 
quantum states 
arise as covariance operators of subquantum random fields. And, although finally at the level of measurement we operate
with their normalized versions, i.e., the standard quantum states, on the subquantum level  the magnitude of the trace plays 
an important role.}

The natural question arises: {\it Can probabilities for a (novel) generalized quantum observable be always represented as probabilities
with respect to some POVM?} 

By taking into account the above discussion on a possibility to express probabilities with respect to a generalized quantum state
as probabilities with respect to the corresponding standard quantum state we can formulate this question in the following way.
Denote the space of all density operators by the symbol ${\cal D}$ and the space of all probability measures on  
$(X,{\cal F})$ by the symbol ${\cal M}.$ Then each POVM, $B \to W(B), B \in {\cal F},$ induces the map 
\begin{equation}
\label{MAP}
j_W: {\cal D} \to {\cal M}, p_W(B\vert \rho)\equiv j_W(\rho)(B)= \rm{Tr} \rho W(B).
\end{equation}  
Each (novel) generalized quantum observable, $B \to E(B),   
B \in {\cal F},$ induces the map 
\begin{equation}
\label{MAP1}
i_E: {\cal D} \to {\cal M}, p_E(B\vert \rho)\equiv j_E(\rho)(B)= \frac{\rm{Tr} \rho \;E(B)}{\rm{Tr}
\rho \;E(X)}.
\end{equation}  
We are interested whether each map $i_E$ can be represented as the  $j_W$-map for some POVM $W.$ 

The answer is negative which is demonstrated by the following example.

{\bf Example.} Consider a generalized observable acting in the qubit space and given by  
two operators: $E_0= 2 \vert 0\rangle \langle 0\vert$ and $E_1= \vert 1\rangle \langle 1\vert.$
Both operators are Hermitian and positive, but $E_0+ E_1 \not= I.$ Hence, the family $E=\{E_0, E_1\}$ 
is a generalized observable, but not POVM. We have 
$
p_E( 0\vert \rho)= \frac{2 \rho_{00}}{2 \rho_{00} + \rho_{11} }, \; p_E( 1 \vert \rho)= 
\frac{\rho_{11}}{2 \rho_{00} + \rho_{11} }. 
$
Suppose now that there exists  POVM $W=\{W_0, W_1=I-W_0 \}$ such that  
$p_E( \alpha\vert \rho) = p_W( \alpha\vert \rho), \alpha=0,1.$
Consider matrix of  $W_0$ in the basis $\vert 0\rangle,\vert 1\rangle:$
$W_0=(x_{ij}).$  Take 
three density matrices $\rho_1=\rm{diag}(1,0), \rho_2=\rm{diag}(0,1), \rho_3=\rm{diag}(1/2,1/2).$ Then
$p_E(0 \vert \rho_1)=1, p_E(0 \vert \rho_2)=0, p_E(0 \vert \rho_3)=2/3.$
Hence, $p_W(0 \vert \rho_1)=1= x_{00}, p_W(0 \vert \rho_2)=0=x_{11}, p_W(0 \vert \rho_3)=2/3= (x_{00}+ x_{11})/2= 1/2.$  
       
The main difference of the generalized quantum probabilities from the conventional quantum probabilities 
is that the new model is {\it nonlinear} with respect to the density operator and the conventional model 
is linear. Hence, it seems that quantum-like applications in cognitive science and psychology lead to nonlinear
calculus of probabilities generalizing the quantum probabilistic calculus.

\end{document}